\documentclass[aps,pra,preprint,showpacs,amsmath,amssymb]{revtex4}
\usepackage{bm}
\usepackage{hyperref}
\renewcommand{\vec}[1]{\bm{#1}}
\renewcommand{\asymp}{\sim}
\renewcommand{\triangleq}{=}

\begin{document}

\title{Asymptotic phase shifts and Levinson theorem for 2D potentials with inverse square singularities}

\author{Denis D. Sheka}
\affiliation{National Taras Shevchenko University of Kiev, 03127 Kiev, Ukraine}
\email{Denis_Sheka@univ.kiev.ua} %
\homepage{http://users.univ.kiev.ua/~Denis_Sheka/index.html} %
\author{Boris A. Ivanov}
\affiliation{Institute of Magnetism, NASU, 03142 Kiev, Ukraine}
\author{Franz G. Mertens}
\affiliation{Physikalisches Institut, Universit\"at Bayreuth, D--95440 Bayreuth,
Germany}

\date{November 11, 2002}

\begin{abstract}
The Levinson theorem for two--dimensional scattering is generalized for
potentials with inverse square singularities. By this theorem, the number of bound states
in a given $m$--th partial wave is related to the phase shift and the singularity strength
of the potential. For the $m$--wave phase shift the asymptotic behaviour is calculated for
short wavelengths.
\end{abstract}


\maketitle


\section{Introduction \label{sec:introduction}}

The Levinson theorem sets up a relation between the number of bound (b) states
$N_l^{\text{b}}$ in a given $l$-th partial wave and  the phase shift $\delta_l(k)$.
The theorem was proved for three-dimensional (3D) central potentials $V(|\vec{r}|)$,
see the review \cite{Taylor72,Newton82}. Levinson's relation for
the $l$--wave phase shift gives $\delta_l(0)-\delta_l(\infty) = \pi N_l^{\text{b}}$.
If the half--bound (hb) state occurs for the $s$-wave type ($l=0$), this is modified to
$\delta_0(0)-\delta_0(\infty) = \pi (N_0^{\text{b}}+\frac12)$. The
Levinson theorem is one of the most beautiful results of scattering theory; it
was a subject of studies by many authors.

The Levinson theorem in 3D has been discussed for noncentral potentials
\cite{Newton60,Newton82,Rosenberg99}, singular potentials \cite{Swan63},
energy--dependent potentials \cite{Sparenberg00a}, nonlocal interactions
\cite{Weber99}, Dirac particles \cite{Ma85,Polyatzky93}, systems with coupling
\cite{Kiers96}, multichannel scattering \cite{Vidal92,Rosenberg98}, multiparticle
single--channel scattering \cite{Rosenberg96}, and in the inverse scattering theory, even
with singular potentials \cite{Sparenberg97,Sparenberg00,Samsonov02}.

Recently, the Levinson theorem was established for lower--dimensional systems,
which play an important role in modern physics of condensed matter and in field theories.
The 1D Levinson theorem was validated for the Schr\"{o}dinger equation \cite{Dong00a},
the Schr\"{o}dinger equation with a nonlocal interaction \cite{Dong00b}, the Klein--Gordon
equation \cite{Dong00c}, and the Dirac equation \cite{Lin99}, even in the presence of
solitons \cite{Gousheh02}. The Levinson theorem was implemented in the (1+1) gauge theory
to calculate the fractional and integer fermion numbers \cite{Farhi01}.

Let us consider 2D systems. The 2D Levinson theorem was established
for different models, too: for the Schr\"{o}dinger equation \cite{Lin97,Dong98},
the Klein--Gordon equation \cite{Dong99}, and the Dirac equation \cite{Lin98}.
Moreover there exists an extension of the Levinson theorem for the Schr\"{o}dinger
equation in D dimensions. There are several methods for studying  the lower-dimensional
Levinson theorem: the Jost function method \cite{Barton85}, the Green function method
\cite{Bolle86,Lin97,Lin98,Lin99}, and the Sturm--Liouville theorem
\cite{Dong98,Dong99,Dong00a,Dong00b,Dong00c,Dong02}.
Levinson's relation for the partial wave phase has the usual form, as for the 3D case;
but the half--bound state for the $p$--wave ($l=1$) contributes exactly like the bound state
and gives an additional $\pi$ to Levinson's relation \cite{Bolle86}.
Let us remind that a half--bound state is the zero--energy solution for the case when the
eigenfunction is finite, but does not decay fast enough at infinity to be square integrable.
In the 2D case a possible $s$--wave half--bound state does not contribute at all to Levinson's
relation, but only the $p$--wave half--bound state. An experimental justification of the
Levinson theorem in the 2D case was made in Refs.~\cite{Portnoi97,Portnoi98} for the 2D plasma.
All mentioned papers, which discuss the 2D version of the Levinson theorem, consider potentials
which are less singular than $|\vec{r}|^{-2}$. This is a standard assumption, which results in the
above mentioned form of the Levinson theorem.

At present singular potentials become an object of interest. Singular
potentials naturally appear in singular inverse problems, i.e. in a supersymmetric
approach to the inverse scattering in 3D, when bound states are removed from the
regular potential \cite{Sparenberg97,Sparenberg00,Samsonov02}. The short distance behaviour of
the singular potential is defined by the inverse square asymptotics at the origin
$V(r)\asymp \beta_0/r^2$; therefore the resulting effective potential for the partial wave $U_l$
(\emph{partial potential}) in the 3D case has the asymptotic form
\begin{equation*}
U_l(r)\triangleq V(r) + \frac{l(l+1)}{r^2} \underset{r\to0}{\asymp}
\frac{\nu(\nu+1)}{r^2},
\end{equation*}
with the \emph{singularity strength} $\nu=\sqrt{(l+1/2)^2+\beta_0}-1/2\neq l$. One can see that the singular
potential acts as a correction to the centrifugal barrier $l(l+1)/r^2$.  The scattering problem for
such potentials with an inverse square singularity was solved firstly by Swan, who has
generalized the Levinson theorem for singular potentials in the 3D case \cite{Swan63}. It reads:
\begin{equation} \label{eq:Swan}
\delta_l(0)-\delta_l(\infty) = \pi\cdot \left(N_l^{\text{b}} + \frac{\nu-l}{2}\right).
\end{equation}
In addition to the general importance for the scattering theory, the generalized
Levinson theorem \eqref{eq:Swan} is useful for the inverse scattering theory, because it
gives a possibility to determine the parameter of the singular core of the potential
from the scattering data.

In the present paper we establish the 2D analogue of the generalized Levinson
theorem \eqref{eq:Swan}. Singular potentials appear in different 2D systems: in
the (2+1)-dimensional O(3)--models like $3D-SU(N_f)$ skyrmions in $N_f$--flavor
meson fields \cite{Walliser00}; in the $2D-O(3)$ spin textures as charged
quasi--particles in ferromagnetic quantum Hall systems \cite{Prange90}; in
different models of 2D magnets as an effective potential of soliton
(vortex)--magnon interaction \cite{Ivanov98,Ivanov99,Sheka01,Ivanov02}.

The paper is organized as follows. In Sec.~\ref{sec:model} we formulate the
scattering problem in the 2D case. We discuss the possible supersymmetric nature of
singular potentials. The scattering problem is solved for the simplest example of a
singular potential, i.e. for the centrifugal model, in Sec.~\ref{sec:centrifugal}.
A simple qualitative picture of the scattering problem is discussed in
Sec.~\ref{sec:WKB}. In this section we calculate the phase shift in the
short--wavelength limit. The generalized Levinson theorem is proved in
Sec.~\ref{sec:proof}. A discussion and concluding remarks are presented in
Sec.~\ref{sec:conclusion}.


\section{Scattering in two dimensions: notations, partial wave method,
singular potentials \label{sec:model}}

Let us consider the Schr\"{o}dinger--like equation in two dimensions:
\begin{equation} \label{eq:Schroedinger}
-\nabla^2\Psi+V(\vec{r})\Psi=i\partial_t\Psi.
\end{equation}

For the central (axially symmetric) potentials, $V(\vec{r})=V(\rho)$,  we apply the
standard partial wave expansion, using the \emph{ansatz}
\begin{equation} \label{eq:ansatz}
\Psi(\vec{r},t)=\sum_{m=-\infty}^\infty\psi_m^{\mathcal{E}}(\rho)\cdot
\exp(im\chi-i\mathcal{E}t)\;,
\end{equation}
where $(\rho,\chi)$ are the polar coordinates in two spatial dimensions,
$\{m,\mathcal{E}\}$ the complete set of eigennumbers, $\mathcal{E}$ and $m$ the energy
and the azimuthal quantum number, respectively. Each partial wave
$\psi_m^{\mathcal{E}}$ is an eigenfunction of the spectral problem
\begin{subequations} \label{eq:EVP&Um}
\begin{equation} \label{eq:EVP}
H\psi_m^{\mathcal{E}}(\rho) = \mathcal{E}\psi_m^{\mathcal{E}}(\rho)
\end{equation}
for the 2D radial Schr\"{o}dinger operator $H=-\nabla_\rho^2+U_m(\rho)$ with the
partial potential
\begin{equation} \label{eq:Um}
U_m(\rho) \triangleq V(\rho)+\frac{m^2}{\rho^2}\;.
\end{equation}
\end{subequations}

Let us formulate the scattering problem. The continuum spectrum exists for
$\mathcal{E}>0$. Note that the eigenfunctions for the free particle, $V(\rho)=0$, have
the form
\begin{equation} \label{eq:psi-free}
\psi_m^{\text{free}}(\rho)\propto J_m(k\rho), \qquad
k\triangleq\sqrt{\mathcal{E}}>0\;,
\end{equation}
where $k$ is a ``radial wave number'', and $J_m$ is a Bessel function. The free
eigenfunctions like $\psi_m^{\text{free}}$ play the role of partial cylinder
waves of the plane wave
\begin{equation} \label{eq:plane-wave}
\exp(i\vec{k}\cdot\vec{r}-i\mathcal{E} t) = \sum_{m=-\infty}^\infty i^m
J_m(k\rho)e^{im\chi-i\mathcal{E}t}.
\end{equation}

The behaviour of the eigenfunctions in the potential $V(\rho)$ can be analyzed at
large distances from the origin, $\rho\gg R$, where $R$ is a typical range of the
potential $V(\rho)$. In view of the asymptotic behaviour $U_m(\rho)\asymp m^2/\rho^2$,
which is valid for fast decreasing potentials $V(\rho)$, in the leading approximation
in $1/\rho$ we have the usual result
\begin{subequations} \label{eq:psi-scattering}
\begin{equation} \label{eq:psi-scat}
\psi_m^{\mathcal{E}} \propto J_{|m|}(k\rho) + \sigma_m(k)Y_{|m|}(k\rho)\;,
\end{equation}
where $Y_m$ is a Neumann function. The quantity $\sigma_m(k)$ stems from the
scattering; it can be interpreted as the scattering amplitude. In the limiting case
$k\rho\gg |m|$ it is convenient to consider the asymptotic form of
Eq.~\eqref{eq:psi-scat},
\begin{equation} \label{eq:psi-scat-as8}
\psi_m^{\mathcal{E}} \propto \frac{1}{\sqrt{\rho}}\cos\left(k\rho - \frac{|m|\pi}{2}
- \frac{\pi}{4}+\delta_m(k)\right),
\end{equation}
\end{subequations}
where the scattering phase, or the phase shift $\delta_m(k)=-\arctan\sigma_m(k)$.
The phase shift contains all informations about the scattering process. In particular,
we give the general solution of the scattering problem for the plane wave
\eqref{eq:plane-wave}. With Eqs.~\eqref{eq:ansatz} and \eqref{eq:psi-scat}, the asymptotic
solution of the Schr\"{o}dinger--like equation \eqref{eq:Schroedinger} for $\rho\gg R$
can be written
\begin{equation} \label{eq:Psi-general}
\begin{split}
\Psi(\vec{r},t)&=\sum_{m=-\infty}^\infty C_m \left(J_{|m|}(k\rho) +
\sigma_m(k)Y_{|m|}(k\rho)\right)\\
&\times \exp(im\chi-i\mathcal{E}t)\;,
\end{split}
\end{equation}
where $C_m$ are constants. To solve the scattering problem for the plane wave let us
choose the constants $C_m$ by comparing Eq.~\eqref{eq:Psi-general} with the
expansion \eqref{eq:psi-free} for the free motion.  Using the asymptotic forms for
the cylinder functions in the region $\rho\gg 1/k$, we obtain
\begin{equation} \label{eq:Psi-via-f}
\begin{split}
\Psi(\vec{r},t) &=e^{i\vec{k}\cdot\vec{r}-i\mathcal{E}t} +
\mathcal{F}(\chi)\frac{e^{ik\rho-i\mathcal{E}t}}{\sqrt{\rho}},\\
\mathcal{F}(\chi) &= \frac{\exp(-i\pi/4)}{\sqrt{2\pi
k}}\cdot\sum_{m=-\infty}^{\infty} \left(e^{2i\delta_m}-1\right)\cdot e^{im\chi}.
\end{split}
\end{equation}
The total scattering cross section is given by the expression
\begin{equation*}
\varrho = \int_0^{2\pi}\!|\mathcal{F}|^2d\chi = \sum_{m=-\infty}^\infty\!
\varrho_m\;,
\end{equation*}
where $\varrho_m = (4/k)\sin^2\delta_m$ are the partial scattering cross sections.

For regular 2D potentials $V(\rho)$, the 2D analogue of the Levinson theorem has
the form \cite{Bolle86,Lin97,Dong98}
\begin{equation} \label{eq:Levinson}
\delta_m(0)-\delta_m(\infty) = \pi\cdot \left(
N_m^{\text{b}}+N_m^{\text{hb}}\cdot\delta_{|m|,1}\right).
\end{equation}
Here the potential $V(\rho)$ satisfies the asymptotic conditions
\begin{subequations} \label{eq:as08}
\begin{eqnarray} \label{eq:as0}
\lim_{\rho=0}\rho^2V(\rho)&=&0,\\ %
\label{eq:as8} %
\lim_{\rho=\infty}\rho^2V(\rho)&=&0,
\end{eqnarray}
\end{subequations}
which provide a regular behaviour at the origin, and fast decaying at infinity.

Now we switch to the singular potentials, having in mind to reestablish the Levinson
theorem.


\subsection{Potentials with inverse square singularity \label{sec:singular-potentials}}

Let us consider potentials with inverse square singularity. At the origin, the
potential has an asymptotics like $V(\rho)\asymp \beta_0/\rho^2$; the corresponding
partial potential \eqref{eq:Um}
\begin{equation} \label{eq:U-nu}
U_m(\rho)\underset{\rho\to0}{\asymp} \frac{\nu^2}{\rho^2}\;, \quad \text{with}\quad
\nu=\sqrt{m^2+\beta_0}\neq m\;.
\end{equation}
Singular potentials like \eqref{eq:U-nu} appear in various 2D non--linear field
theories, e.g. for the scattering problem of linear excitations by
topological solitons \cite{Walliser00,Prange90,Ivanov98,Ivanov99,Sheka01}.

Moreover singular potentials naturally appear from regular ones under Darboux
transformations \cite{Sparenberg97,Sparenberg00,Leeb00,Samsonov02}. Let us recall the principle of
Darboux (supersymmetric) transformations for the 2D case \cite{Ivanov99}. We suppose
that the spectral problem \eqref{eq:EVP} has at least one bound state
$\mathcal{E}_0<0$. Assuming that we start from the regular potential under
conditions \eqref{eq:as08}, then the eigenfunction
$\psi_0\equiv\psi_m^{\mathcal{E}_0}(\rho)$ may have the following asymptotic
behaviour
\begin{equation} \label{eq:psi-as}
\psi_0(\rho)\propto
  \begin{cases}
    \rho^{|m|}, & \text{when $\rho\to0$}\;, \\
    \rho^{-1/2}\cdot\exp\left(-\kappa\rho\right),
               & \text{when $\rho\to\infty$}\;,
  \end{cases}
\end{equation}
where $\kappa = \sqrt{-\mathcal{E}_0}>0$.

To explain the method we introduce the Hermitian--conjugate lowering and raising
operators \cite{Ivanov99}
\begin{equation} \label{eq:A&A+}
A = -\frac{d}{d\rho}+W(\rho),\qquad A^\dag = \frac{d}{d\rho}+\frac{1}{\rho}+
W(\rho),
\end{equation}
where the superpotential
\begin{equation} \label{eq:superpotential}
W(\rho)\triangleq \frac{d}{d\rho}\ln \psi_0
\end{equation}
is such that $A\psi_0=0$. By introducing these operators we can represent the
Schr\"{o}dinger operator $H$ in the factorized form
\begin{equation} \label{eq:H-via-A&A+}
H=A^\dag A + \mathcal{E}_0\;,
\end{equation}
the factorization energy $\mathcal{E}_0$ coincides with the energy of the bound
state. Such a factorization makes it possible to reformulate the initial problem
\eqref{eq:H-via-A&A+} in terms of the eigenfunction $\tilde{\psi}_m\triangleq
A\psi_m$ of the spectral problem
\begin{equation} \label{eq:tilde-H}
\tilde{H}\triangleq AA^\dag + \mathcal{E}_0 = -\nabla_\rho^2 + \tilde{U}_m(\rho)\;,
\end{equation}
where the partial potential
\begin{equation} \label{eq:tilde-Um}
\tilde{U}_m(\rho)\triangleq U_m(\rho)+\frac{1}{\rho^2}-2\frac{d}{d\rho}W(\rho)\;.
\end{equation}
Taking into account the conditions \eqref{eq:psi-as}, one can derive the asymptotic
behaviour of the partial potential $\tilde{U}_m$,
\begin{equation} \label{eq:tilde-U-as}
\tilde{U}_m(\rho)\asymp
  \begin{cases}
\dfrac{\nu^2}{\rho^2}\quad \text{with $\nu=|m|-1$}, & \text{when $\rho\to0$}\;, \\
    \dfrac{m^2}{\rho^2},& \text{when $\rho\to\infty$}\;.
  \end{cases}
\end{equation}
We see that the eigenspectrum of the new spectral problem \eqref{eq:tilde-H} does not
contain the bound state $\psi_0$. The resulting potential has a singularity; in fact,
the partial potential $\tilde{U}_m(\rho)$ corresponds to the particle potential
$V(\rho)=\beta/\rho^2$ with the parameter $\beta=1-2|m|$. After a series of $n$
transformations like \eqref{eq:tilde-H}, we remove $n$ bound states from the
spectrum, which results in $\tilde{U}_m\asymp \nu^2/\rho^2$, with $\nu=|m|-n.$


\subsection{Potentials with inverse square tail \label{sec:tail-potentials}}

Let us discuss potentials with an inverse square tail, when far from the origin the
potential $V(\rho)\asymp \beta_\infty/\rho^2$; the corresponding partial potential
\begin{equation} \label{eq:U-mu}
U_m(\rho)\underset{\rho\to\infty}{\asymp} \frac{\mu^2}{\rho^2}\;, \quad
\text{with}\quad \mu=\sqrt{m^2+\beta_\infty}\neq m\;.
\end{equation}
Potentials like \eqref{eq:U-mu} are of interest in field theories: in the (2+1)
nonlinear $\sigma$--model of the $\vec{n}$--field \cite{Walliser00,Ivanov99}, in models
of 2D easy--axis \cite{Sheka01} and easy--plane ferromagnets in the cone state
\cite{Ivanov02}.

To study the scattering problem let us consider the asymptotic behaviour of the
eigenfunctions. Obviously, at large distances $\rho\gg R$, where the scattering
approximation is valid, one can use the partial wave expansion by the cylinder
functions of the integer indexes only; then the eigenfunction $\psi_m^{\mathcal{E}}$
can be written as $J_{|m|}+\sigma_m Y_{|m|}$ with the asymptotic form
\eqref{eq:psi-scat-as8}.

On the other hand, in the leading approximation in $1/k\rho$, the solution of the
Schr\"odinger equation \eqref{eq:EVP} with the potential \eqref{eq:U-mu} can be
written as
\begin{equation} \label{eq:psi-mu}
\begin{split}
\psi_m^{\mathcal{E}}(\rho)&\propto J_{|\mu|}(k\rho) +
{\tilde{\sigma}}_{\mu}(k)Y_{|\mu|}(k\rho)\\
&\propto \frac{1}{\sqrt{\rho}}\cos\left(k\rho - \frac{|\mu|\pi}{2} -
\frac{\pi}{4}+{\tilde{\delta}}_{\mu}(k)\right),
\end{split}
\end{equation}
where the index of the cylinder functions $\mu\neq m$, see Eq.~\eqref{eq:U-mu}.

The phase shift $\delta_m$ can be calculated from ${\tilde{\delta}}_{\mu}$ by
comparing Eqs.~\eqref{eq:psi-scat-as8} and \eqref{eq:psi-mu},
\begin{equation} \label{eq:delta-via-delta-mu}
\delta_m(k) = {\tilde{\delta}}_{\mu}(k) + \frac{|m|-|\mu|}{2}\cdot\pi\;,
\end{equation}
in accordance with the results of Refs.~\cite{Dong98,Ivanov02}. Note that
Levinson's relation has the same form for both phase shifts $\delta_m$ and
${\tilde{\delta}}_{\mu}$,
\begin{equation*}
\delta_m(0)-\delta_m(\infty) = {\tilde{\delta}}_{\mu}(0)
-{\tilde{\delta}}_{\mu}(\infty)\;.
\end{equation*}


\section{Scattering problem for the centrifugal model \label{sec:centrifugal}}

For the analytical description of the scattering problem, let us consider the
simplest model, which includes the main features of the problem, having both inverse
square singularity and inverse square tail. The partial potential of this very
simple \emph{centrifugal model} \cite{Sheka01} has the form
\begin{equation} \label{eq:U-centrifugal}
U_m^{\text{cf}}(\rho) =
  \begin{cases}
    \dfrac{\nu^2}{\rho^2}\;, & \text{when $\rho<R$}\;, \\
    \dfrac{\mu^2}{\rho^2}\;, & \text{otherwise}\;,
  \end{cases}
\end{equation}
with $\nu\neq m$, and $\mu\neq m$.

This model describes a quasi--free particle in each of the regions $\rho<R$ and
$\rho>R$. The only effect of the interaction with the potential $U_m^{\text{cf}}$ is
a shift of the mode indices:
\begin{equation} \label{eq:psi-cenrifugal}
\psi_m^{\text{cf}}(r) \propto
  \begin{cases}
    J_{|\nu|}(k\rho)\;, & \text{when $\rho<R$}, \\
    J_{|\mu|}(k\rho) + {\tilde{\sigma}}_\mu(k) Y_{|\mu|}(k\rho),
                     & \text{otherwise}\;.
  \end{cases}
\end{equation}
The usual matching condition for these solutions has the form
\begin{equation} \label{m-cond}
\left[\frac{\psi'}{\psi}\right]_R = 0\;,
\end{equation}
where $\left[\dots\right]_R\equiv (\dots)\bigr|_{R+0} - (\dots)\bigr|_{R-0}$, and
the prime denotes $d/d\rho$. The calculations lead to the scattering phase shift in the
form:
\begin{equation} \label{eq:delta-centrifugal}
\begin{split}
\delta_m^{\text{cf}}(k) &= \frac{|m|-|\mu|}{2}\cdot\pi - \arctan
{\tilde{\sigma}}_{\mu}^{\text{cf}}(\varkappa\equiv kR)\;,\\
{\tilde{\sigma}}_{\mu}^{\text{cf}}(\varkappa) &= \frac{
J_{|\nu|}^{\prime}(\varkappa)\cdot J_{|\mu|}(\varkappa)-
J_{|\mu|}^{\prime}(\varkappa)\cdot J_{|\nu|}(\varkappa)} {J_{|\nu|}(\varkappa)\cdot
Y_{|\mu|}^{\prime}(\varkappa)- J_{|\nu|}^{\prime}(\varkappa)\cdot
Y_{|\mu|}(\varkappa)}\;.
\end{split}
\end{equation}

Using the asymptotic form of the cylinder functions, one can find the long-- and
short--wavelength behaviour of the phase shift \eqref{eq:delta-centrifugal},
\begin{equation} \label{eq:delta-centrifugal-as}
\delta_m^{\text{cf}}(k) \asymp
  \begin{cases}
   \dfrac{|m|-|\mu|}{2}\cdot\pi + \mathcal{A}_m \cdot
   \left(\dfrac{kR}{2}\right)^{2|\mu|}\!\!\!\!,
   & kR\ll 1, \\
   \dfrac{|m|-|\nu|}{2}\cdot\pi - \dfrac{\mu^2-\nu^2}{2kR}, & kR\gg 1,
  \end{cases}
\end{equation}
where $\mathcal{A}_m = -\dfrac{\pi|\mu|}{(|\mu|)!^2} \cdot
\dfrac{|\mu|+|\nu|}{|\mu|-|\nu|}$.

The Levinson theorem for the centrifugal model can be easily derived from
Eq.~\eqref{eq:delta-centrifugal-as}:
\begin{equation} \label{eq:Levinson-centrifugal}
\delta_m^{\text{cf}}(0) - \delta_m^{\text{cf}}(\infty) = \pi\cdot
\frac{|\nu|-|\mu|}{2}\;.
\end{equation}


\section{Scattering problem in the WKB approximation \label{sec:WKB}}

Now we discuss the general case where the partial potential has the asymptotic
behaviour
\begin{equation} \label{eq:U-as}
U_m(\rho) \asymp
  \begin{cases}
    \dfrac{\nu^2}{\rho^2}, & \text{when $\rho\to0$}, \\
    \dfrac{\mu^2}{\rho^2}, & \text{when $\rho\to\infty$},
  \end{cases}
\end{equation}
with $\nu\neq m$, and $\mu\neq m$.

The scattering problem can be treated analytically in the short--wavelength limit,
$kR\gg1$. It is natural to suppose that the WKB--approximation is valid for this
case. We use the WKB--method in the form proposed earlier for the description of
the scattering for isotropic 2D magnets \cite{Ivanov99}. We start from the
effective 1D Schr\"odinger equation for the radial function $\psi_m(\rho) =
u_m(\rho)/\sqrt{\rho}$, which yields
\begin{equation} \label{eq:U-ef}
\begin{split}
&\left[-\frac{d^2}{d\rho^2}+\mathcal{U}_{\text{eff}}(\rho)\right] u_m = \mathcal{E} u_m\;,\\
&\mathcal{U}_{\text{eff}}(\rho)\triangleq V(\rho)+\frac{4m^2-1}{4\rho^2}\;.
\end{split}
\end{equation}

The WKB--solution of the Eq.~\eqref{eq:U-ef}, i.e. the 1D wave function $u_m^{WKB}$,
leads to the following form of the partial wave
\begin{equation} \label{eq:WKB}
\psi_m^{WKB} = \frac{u_m^{WKB}}{\sqrt{\rho}} \propto
\frac{1}{\sqrt{\rho\cdot\mathcal{P}(\rho)}}\cos\left(\chi_0 + \int_{\rho_0}^\rho
\mathcal{P}(\rho')d\rho'\right),
\end{equation}
where $\mathcal{P}=\sqrt{k^2-\mathcal{U}_{\text{eff}}}$. The analysis shows that the
Eq.~\eqref{eq:WKB} is valid for $\rho>a$, where $a$ is the turning point. The value
of $a$ corresponds to the condition $\mathcal{P}(a)=0$, which results in $a\sim|m|/k\ll
R$. We assume that the parameter $\rho_0$ satisfies the condition $a\ll \rho_0 \ll R$, hence
$1\sim ka\ll k\rho_0 \ll kR$.

On the other hand, at small distances $\rho\ll R$, the partial potential
$U_m\asymp\nu^2/\rho^2$, i.e. it describes the free particle in the form
\eqref{eq:psi-free} with a shifted index:
\begin{equation} \label{eq:psi-WKB0}
\psi_m^{\mathcal{E}}(\rho) \propto  J_{|\nu|}(k\rho),\qquad\text{when $\rho\ll R$}.
\end{equation}
For $kR\gg |\nu|$ there is a wide range of values of $\rho$, namely
\begin{equation} \label{eq:wide-range}
|\nu|/k\ll \rho\ll R,
\end{equation}
where we can use the asymptotic expression  for the Bessel function
\eqref{eq:psi-WKB0} in the limit $k\rho\gg|\nu|$ \cite{Ivanov99}:
\begin{equation} \label{eq:psi-WKB4rho>>1}
\psi_m^{\mathcal{E}}(\rho) \propto  \frac{1}{\sqrt{\rho}}\cos\Bigl(k\rho -
\frac{|\nu|\pi}{2} -\frac\pi4+\frac{4\nu^2-1}{8k\rho}\Bigr).
\end{equation}
In the range \eqref{eq:wide-range} the solutions \eqref{eq:WKB} and
\eqref{eq:psi-WKB4rho>>1} agree due to an overlap in the entire range of parameters,
so one can derive the phase $\chi_0$ in the WKB--solution \eqref{eq:WKB},
\begin{equation*}
\chi_0 = k\rho_0 - \frac{|\nu|\pi}{2} -\frac\pi4+\frac{4\nu^2-1}{8k\rho_0}\;.
\end{equation*}
Therefore, we are able to calculate the short--wavelength asymptotic expression for
the scattering wave phase shift by the asymptotic expansion of the WKB--solution
\eqref{eq:WKB}:
\begin{equation} \label{eq:delta-via-P}
\begin{split}
\delta_m(k) = \lim_{\rho\to\infty}&
\Biggl(\int_{\rho_0}^\rho\mathcal{P}(\rho')d\rho' +\chi_0 -
k\rho\\
&+\frac{|m|\pi}{2}+\frac{\pi}{4}-\frac{4m^2-1}{8k\rho}\Biggr).
\end{split}
\end{equation}
Under the condition $k\rho\gg1$, the WKB--integral in \eqref{eq:delta-via-P}
can be calculated in the leading approximation in $1/k\rho$,
\begin{equation*}
\int_{\rho_0}^\rho\mathcal{P}(\rho')d\rho' \approx k(\rho-\rho_0)-\frac{1}{2k}
\int_{\rho_0}^\rho \mathcal{U}_{\text{eff}}(\rho')d\rho'.
\end{equation*}
As result, the scattering phase shift for large wave numbers, $k\gg 1/R$, has the form
\begin{equation} \label{eq:delta-WKB}
\delta_m(k) = \pi\cdot\frac{|m|-|\nu|}{2} - \frac{1}{2k}\int_0^\infty\Delta
U_m(\rho')d\rho',
\end{equation}
which contains a general, so--called eikonal dependence $\delta\propto 1/k$. The
potential $\Delta U_m$,
\begin{equation*}
\Delta U_m(\rho) \triangleq U_m(\rho)-\frac{\nu^2}{\rho^2} =
V(\rho)+\frac{m^2-\nu^2}{\rho^2}
\end{equation*}
has no singularities at the origin, $\lim\limits_{\rho=0}\rho^2\Delta U_m(\rho)=0$.
Note that the scattering phase shift does not tend to zero for $k\to\infty$, but at
some finite value $(\pi/2)\cdot\left(|m|-|\nu|\right)$. This feature is caused by
the singularity of the potential at the origin.

Let us discuss the Levinson theorem. Before proceeding to a formal proof (see the next
section), it will be useful to present a heuristic argument. We consider the
scattering problem near the threshold, $k=0$. Let us suppose that the potential well
is so deep that bound states for $\mathcal{E}\lesssim0$ can be described by the WKB
approximation \eqref{eq:WKB} with the phase shifts given by \eqref{eq:delta-via-P}.
The WKB--integral in \eqref{eq:delta-via-P} can be calculated in the leading
approximation in $k\rho$,
\begin{equation} \label{eq:Bohr-Sommerfeld(1)}
\int_{\rho_0}^\rho\mathcal{P}(\rho')d\rho' \approx \int_a^b\mathcal{P}(\rho')
d\rho'+ k\rho+\text{const},
\end{equation}
where $a$ and $b$ are the turning points of the quasiclassical motion in the
potential $\mathcal{U}_{\text{ef}}$, see Eq.~\eqref{eq:U-ef}. Under such assumptions
the Bohr--Sommerfeld quantization rule is valid,
\begin{equation} \label{eq:Bohr-Sommerfeld(2)}
\int_a^b \mathcal{P}(\rho')d\rho'=\pi\cdot\left(N_m^{\text{b}} + \gamma\right),
\end{equation}
where $\gamma$ depends on the potential's behaviour near the turning points. Note that
the bound states are absent, $N_m^{\text{b}}=0$, for the limiting case of the shallow
well, $V\to0$; this is the case for the Born approximation with the general
scattering condition $\delta_m(0)=0$ \cite{Taylor72,Newton82}. Using Eqs.
\eqref{eq:Bohr-Sommerfeld(1)}, \eqref{eq:Bohr-Sommerfeld(2)}, this results in the phase
shift \eqref{eq:delta-via-P} in the form
\begin{equation*}
\delta_m(0)=\pi \cdot N_m^{\text{b}}
\end{equation*}
for regular potentials. However, in the case of potentials with inverse square
singularity, it should be shifted by $(\pi/2)\cdot(|m|-|\mu|)$, see
Eq.~\eqref{eq:delta-via-delta-mu}. Thus, our simple qualitative picture leads to the
long--wavelength limit of the phase shift,
\begin{equation} \label{eq:delta(0)-WKB}
\delta_m(0)= \pi\cdot \left( N_m^{\text{b}}+ \frac{|m|-|\mu|}{2}\right).
\end{equation}

Using the limiting values for the phase shift, Eqs.~\eqref{eq:delta-WKB},
\eqref{eq:delta(0)-WKB}, one can calculate Levinson's relation:
\begin{equation} \label{eq:Levinson-WKB}
\delta_m(0) - \delta_m(\infty) = \pi\cdot \left( N_m^{\text{b}}+
\frac{|\nu|-|\mu|}{2}\right).
\end{equation}


\section{The Levinson theorem \label{sec:proof}}

Let us enter into a proof of the Levinson theorem. There are three main methods to
derive the theorem: the Jost functions method, the Green's functions method, and
the Sturm--Liouville method, which were used for the 3D case, for the details see
Ref.~\cite{Dong98}.

To generalize the Levinson theorem, we use the method of the Green functions, as it
was done for regular potentials by \citet{Lin97}. We consider the noncritical case,
when the Schr\"{o}dinger equation has no half bound states.

The idea of Lin's method \cite{Lin97} is to count the number of states in the
system by two different ways.

The continuous part of the spectrum is discretized to count the number of scattering
states. Therefore, the total (infinite) number of states in the system does not
depend on the shape of the potential, it results in
\begin{subequations} \label{eq:G-G0}
\begin{equation} \label{eq:G-G0-total}
\text{Im}\!\!\int_{-\infty}^\infty\!\! d\mathcal{E}\!\!\int_0^\infty \!\! \rho d\rho
\left\{ G[U_m] - G[U_m^{\text{free}} ] \right\} = 0,
\end{equation}
where $G[U_m]\equiv G_m(\rho,\rho,\mathcal{E};U_m)$ and $G[U_m^{\text{free}}]\equiv
G_m(\rho,\rho,\mathcal{E};U_m^{\text{free}})$ are the Green functions with and
without potential, respectively; and the retarded Green function is defined by
\begin{equation*}
G_m(\rho,\rho',\mathcal{E};U_m) \triangleq \sum_{\kappa} \frac{\psi_m^{\mathcal{E}}
(\rho) \psi_m^{\mathcal{E}}(\rho')}{\mathcal{E}-\mathcal{E}_{m\kappa}+i\epsilon}\;.
\end{equation*}
In this method, the number of bound states,
\begin{equation} \label{eq:G-G0-bound}
\pi N_m^{\text{b}} = -\text{Im}\!\!\int_{-\infty}^0\!\!
d\mathcal{E}\!\!\int_0^\infty \!\! \rho d\rho \left\{ G[U_m] - G[U_m^{\text{free}} ]
\right\}.
\end{equation}
On the other hand, the continuous part of the expression \eqref{eq:G-G0-total} can be
directly calculated without discretization:
\begin{equation} \label{eq:G-G0-scat}
\begin{split}
\text{Im}\!\!\int_{0}^\infty\!\! d\mathcal{E}\!\!\int_0^\infty \!\! \rho d\rho
\left\{ G[U_m] - G[U_m^{\text{free}}]\right\} = \delta_m(0) - \delta_m(\infty).
\end{split}
\end{equation}
\end{subequations}
Combining Eqs.~\eqref{eq:G-G0}, one can obtain the Levinson theorem in the form
\eqref{eq:Levinson}. However, the method of the Green functions in the form proposed by
\citet{Lin97} does not work for singular potentials. The reason is that the difference
of Green functions $G[U_m] - G[U_m^{\text{free}}]$ in \eqref{eq:G-G0} has a singularity at
the origin, hence it is not integrable.

That is why we need to generalize the method for the case of singular potentials.
The idea is to compare the required partial potential $U_m$ not with the free
particle partial potential $U_m^{\text{free}}$, but with another potential $U_m^\star$,
which could compensate the singularities of $U_m$. As we have mentioned before, the
number of states does not depend on the shape of the potential. It means that repeating the same
proof, Eqs.~\eqref{eq:G-G0} can be easily generalized for the systems $G[U_m]$ and
$G[U_m^\star]$ with two different potentials $U_m$ and $U_m^\star$:
\begin{subequations} \label{eq:G4V&V*}
\begin{eqnarray} \label{G4V&V*-total}
&&\text{Im}\!\!\int_{-\infty}^\infty\!\! d\mathcal{E}\!\!\int_0^\infty \!\!
\rho d\rho \left\{ G[U_m] - G[U_m^\star]\right\} = 0,\\
&&\text{Im}\!\!\int_{-\infty}^0\!\! d\mathcal{E}\!\!\int_0^\infty \!\! \rho
d\rho \left\{ G[U_m] - G[U_m^\star]\right\}\nonumber\\
\label{G4V&V*-bound} %
&&=-\pi\cdot\left( N_m^{\text{b}}-N_m^{\text{b}\star}\right),\\
&&\text{Im}\!\!\int_{0}^\infty\!\! d\mathcal{E}\!\!\int_0^\infty \!\! \rho
d\rho \left\{ G[U_m] - G[U_m^\star]\right\}\nonumber\\
\label{G4V&V*-scat} %
&&= \delta_m(0) - \delta_m(\infty) - \delta_m^\star(0) + \delta_m^\star(\infty),
\end{eqnarray}
\end{subequations}
where $N_m^{\text{b}\star}$ and $\delta_m^\star(k)$ are the number of bound states
and the scattering phase shift for the system with the partial potential
$U_m^\star=V^\star+m^2/\rho^2$.

Note that choosing $V^\star=0$, one can obtain Levinson's relation for the regular
potentials in the form of \citet{Lin97}, see Eqs.~\eqref{eq:G-G0}, which leads to the
Levinson theorem \eqref{eq:Levinson}.

However, in the case of a singular potential, we need to choose $V^\star$ in the form
which has the same singularities as the potential $V$. To solve the problem we set
$U_m^\star = U_m^{\text{cf}}$~; hence both partial potentials $U_m$ and
the centrifugal potential $U_m^{\text{cf}}$ have the same features. Therefore,
Eqs.~\eqref{eq:G-G0} with account of Levinson's relation \eqref{eq:Levinson-centrifugal},
lead to the following form:
\begin{equation} \label{eq:Levinson-WKB'}
\tag{\ref{eq:Levinson-WKB}$'$} %
\delta_m(0) - \delta_m(\infty) = \pi\cdot \left( N_m^{\text{b}}+
\frac{|\nu|-|\mu|}{2}\right),
\end{equation}
so we reestablish the generalized Levinson theorem in the form \eqref{eq:Levinson-WKB}.


Let us discuss the result. To explain the meaning of the extra term $(\pi/2)\cdot(|\nu|-|\mu|)$
in the generalized Levinson relation \eqref{eq:Levinson-WKB'}, let us remind
that in the partial wave method the scattering data are classified by the azimuthal quantum
number $m$, which is the strength of the centrifugal potential. In the presence of
the potential with an inverse square singularity at the origin like $U_m\asymp\nu^2/\rho^2$,
the effective singularity strength is shifted by the value $|\nu|-|m|$, which results
in a change in the short--wavelength scattering phase shift by $(\pi/2)\cdot(|m|-|\nu|)$.
The same situation takes place for the potentials with an inverse square tail at infinity
like $U_m\asymp\mu^2/\rho^2$. The effective singularity strength is shifted now by the value
$|\mu|-|m|$, and the long--wavelength scattering data are changed by
$(\pi/2)\cdot(|m|-|\mu|)$. As result, the correction to the Levinson's relation is
\begin{equation*}
\pi\cdot\frac{|m|-|\mu|}{2} - \pi\cdot\frac{|m|-|\nu|}{2} = \pi\cdot\frac{|\nu|-|\mu|}{2}.
\end{equation*}
Such a correction looks like a modification in the classification of the scattered states, both
at the origin ($\psi_m\to\psi_\nu$), and at the infinity ($\psi_m\to\psi_\mu$).
However, we need to stress that the singularity strengths $\nu$ and $\mu$ can assume any real
values, while the quantum number $m$ is always integer.


\section{Conclusion \label{sec:conclusion}}

In conclusion, we have established the analogue of the Levinson theorem in the case
of two--dimensional scattering for central potentials, which are independent
of both the energy and the azimuthal momentum $m$, but have inverse square
singularities and tails.

The presence of $m$--dependent potentials can essentially
change the scattering picture: the symmetry $\delta_m(k)=\delta_{-m}(k)$
is broken, so it is not enough to take into account partial waves with $m\geq0$ only.
As result Levinson's relation \eqref{eq:Levinson-WKB} has a different form for opposite $m$.
Moreover, the threshold behaviour for the half--bound states changes, so
the contribution of the half--bound states in the form \eqref{eq:Levinson} may be not adequate.

The generalized Levinson theorem \eqref{eq:Levinson-WKB'} can be applied to different
physical problems. For example, it becomes a central point in the singular inverse method
\cite{Sparenberg00}, giving a possibility to derive the potential from the
scattering phase shift. At the same time it provides a method to count bound states.
The method can be used in various 2D field theories with applications to the physics of
2D plasma \cite{Portnoi97,Portnoi98}, nuclear physics \cite{Walliser00},
quantum Hall effect \cite{Prange90}, and 2D magnetism
\cite{Ivanov98,Ivanov99,Sheka01,Ivanov02}.

The method of the 2D radial Darboux transformations, considered in the paper, can be
applied to the supersymmetric quantum mechanics, e.g. for the problem of
phase--equivalent potentials \cite{Sparenberg96,Sparenberg97,Sparenberg00,Leeb00,Samsonov02},
even for energy--dependent potentials \cite{Sparenberg00a}.

\begin{acknowledgments}
D.D.Sh. thanks the University of Bayreuth, where part of this work was performed, for
kind hospitality and acknowledges support by the European Graduate School
``Non--equilibrium phenomena and phase transitions in complex systems''.
\end{acknowledgments}




\end{document}